\documentclass[aps,prx,twocolumn,showpacs]{revtex4-1}

\usepackage{inputenc}
\usepackage{graphics}
\usepackage{graphicx}
\usepackage[mathscr]{euscript}
\usepackage{mathrsfs}

\usepackage[normalem]{ulem}
\usepackage{soul}

\ifx\pdftexversion\undefined
\usepackage[dvips]{hyperref}
\else
\usepackage{hyperref}
\fi
\hypersetup{
  colorlinks = true, linkcolor = blue
}

\begin{document}
\preprint{}

\title{Momentum-space atom correlations in a Mott insulator}

\author{C\'ecile Carcy, Hugo Cayla, Antoine Tenart, Alain Aspect, Marco Mancini, and David Cl\'ement}

\email[Corresponding author: ]{david.clement@institutoptique.fr}

\affiliation{Laboratoire Charles Fabry, Institut d'Optique Graduate School, CNRS, Universit\'e Paris Saclay, 2 Avenue Augustin Fresnel, 91127 Palaiseau cedex, France}

\date{\today}

\begin{abstract}
We report on the investigation of the three-dimensional single-atom-resolved distributions of bosonic Mott insulators in momentum-space. Firstly, we measure the two-body and three-body correlations deep in the Mott regime, finding a perfectly contrasted bunching whose periodicity reproduces the reciprocal lattice. In addition, we show that the two-body correlation length is inversely proportional to the in-trap size of the Mott state with a pre-factor in agreement with the prediction for an incoherent state occupying a uniformly filled lattice. Our findings indicate that the momentum-space correlations of a Mott insulator at small tunnelling is that of a many-body ground-state with Gaussian statistics. Secondly, in the Mott insulating regime with increasing tunnelling, we extract the spectral weight of the quasi-particles from the momentum density profiles. On approaching the transition towards a superfluid, the momentum spread of the spectral weight is found to decrease as a result of the increased mobility of the quasi-particles. While the shapes of the observed spectral weight agree with those predicted by perturbative many-body calculations for homogeneous systems, the fitted mobilities are larger than the theoretical ones, mostly because of the co-existence of various phases in the trap.
\end{abstract} 

\maketitle 

\section{INTRODUCTION}

Measuring many-body correlations is central to investigate and reveal the properties of strongly interacting quantum matter \cite{cirac2012}. It allows one to test the predictions of simple microscopic models used to understand many-body physics \cite{mahan2000}. One paradigmatic example is the Hubbard hamiltonian that describes quantum particles hopping between the sites of a lattice in the presence of interaction. Comparison of the predictions of the Hubbard model with the measured response functions, {\it e.g.}, the one-particle spectral function \cite{damascelli2003} and the dynamical structure factor \cite{tranquada1995}, showed its relevance for a large variety of systems, from transition-metal oxides \cite{imada1998} and heavy fermions \cite{stewart1984} to high-$T_{c}$ superconductors \cite{lee2006}. A direct probe of correlations between individual particles is yet hardly possible to implement in solids. Thanks to single-atom-resolved detection methods \cite{ott2016}, quantum gases offer an alternative testbed for many-body theories. 

The celebrated Mott transition, a generic metal-to-insulator transition of Hubbard models \cite{mott1949}, was investigated with quantum gas microscopes, illustrating the benefits from probing many-body physics at the single particle level. Position-space two-particle correlations in Mott insulators have indeed yielded unprecedented and quantitative observations, such as the reduction of on-site atom fluctuations \cite{sherson2010, bakr2010}, the string order in low-dimensions \cite{endres2011}, the out-of-equilibrium dynamics \cite{ma2011, cheneau2012} and the anti-ferromagnetic ordering \cite{mazurenko2017, hilker2017}. On the other hand, the investigation of  \textit{momentum-space correlations} in atomic Mott states are scarce, and in particular there had been no implementation of a single-atom-resolved probe of momentum correlations. Momentum-resolved light scattering techniques were implemented to measure spectral functions \cite{stoferle2004, clement2009, fabbri2012}, and the noise correlation analysis of time-of-flight (TOF) absorption images \cite{altman2004} has revealed density correlations in expanding Mott insulators \cite{folling2005, rom2006}. But these probes have not permitted quantitative studies of correlation functions, including their shape and their width, due to limited resolution in momentum,  line-of-sight integration  and lack  of single-particle sensitivity. Experimental signatures of the contribution of particle-hole excitations to the coherence of a Mott insulator were extracted from the visibility \cite{gerbier2005} and the density modulation of TOF distributions \cite{spielman2007}. But a momentum-resolved measure of this contribution when approaching the Mott transition, {\it i.e.} an observation of the modifications of the spectral weight of the quasi-particles in momentum-space, was not possible. Recently, however, it has become possible to measure momentum space correlations, either in one or two-dimensional geometries using optical imaging \cite{bucker2009,fang2016, preiss2018}, or in three-dimensions with the species metastable Helium (He*), which can be detected at the single atom level \cite{hodgman2017, cayla2018}.

In this work, we report on using momentum-space single-atom-resolved distributions \cite{cayla2018} in  He* gaseous Mott insulators to investigate both the average population of the momentum states ({\it i.e.}, the momentum density) and its fluctuations (through two-body and three-body momentum correlations). Deep in the Mott regime, we observe a perfectly contrasted momentum bunching, with second- and third-order correlation functions whose structure reproduces the reciprocal lattice. Moreover, from the shape of the bunching peaks, we determine the  two-particle correlation length in  Mott insulators of various in-trap sizes, finding an excellent agreement with ab-initio calculations. These features show that in the limit of vanishing tunnelling the Mott correlations are driven by a Gaussian density operator for bosons uniformly distributed in the lattice sites. From cuts in the three-dimensional (3D) momentum density,  we also extract the spectral weight $Z(k)$ of the Mott quasi-particles. The evolution of $Z(k)$ when the system approaches the superfluid-to-Mott transition provides a direct signature of the increased spatial coherence of the quasi-particles. At the quantitative level, however, a comparison with perturbative many-body approaches reveals discrepancies which we discuss.

\section{MOMENTUM SPACE CORRELATION FUNCTIONS OF A MOTT STATE:\\ THEORY IN A NUTSHELL}

A Mott insulator is an insulating phase caused by the presence of strong interactions between particles moving in crystalline structures. Hubbard models provide a microscopic description of the Mott physics based on the interplay between  interaction and kinetic energy. When the ratio $U/J$ of the on-site (repulsive) interaction energy $U$ to the hopping energy $J$ between adjacent lattice sites increases, the ground-state of a Hubbard Hamiltonian undergoes a phase transition from a conductor to a Mott insulator. The Fermi Hubbard model is central for the understanding of metal-insulator transitions in solids when electron-electron interaction are important \cite{mott1949, imada1998}. Recently quantum gases have permitted the observation of a Mott transition with bosons \cite{greiner2002} which is described by the Bose-Hubbard Hamiltonian \cite{jaksch1998},
\begin{equation}
H=-J \sum_{\langle l,l' \rangle} \hat{b}^{\dagger}_{l} \hat{b}_{l'} + \frac{U}{2} \sum_{l=1}^{N_{\rm site}}  \hat{n}_{l} (\hat{n}_{l}-1) \label{Eq:BHH}
\end{equation}
where $\hat{b}_{l}$ is the annihilation operator of a particle on site labelled $l$ and $\hat{n}_{l}=\hat{b}^{\dagger}_{l} \hat{b}_{l} $. The critical value for the Mott transition in the 3D Bose-Hubbard Hamiltonian was calculated numerically, and found equal to $(U/J)_{\rm c}=29.3$ \cite{capogrosso2007}. The properties of the Mott state -- such as the amplitude of the gap in the excitation spectrum or the role of particle-hole excitations -- change when varying $U/J$ from deep in the insulating regime $( U/J \gg (U/J)_{\rm c}$) to the value at the phase transition. In a first approximation, these modifications can be viewed as resulting from the contribution of some particle and hole excitations on top of a uniformly filled lattice. In quantum gas experiments, the ratio $U/J$ can be varied continuously by changing the amplitude of the lattice potential, allowing one to investigate the modifications of that many-body ground-state (the Mott Insulator) with a high degree of control \cite{greiner2002}. In the following we describe the properties of the Mott state in the momentum-space. To this aim we introduce the momentum-space operators 
\begin{equation}
\hat{a}(\vec{k})= \frac{1}{\sqrt{V}} \sum_{l=1}^{N_{\rm site}}  e^{i\vec{k}.\vec{r}_l } \hat{b}_{l} ,\label{Eq:MomOp}
\end{equation} 
where the volume of quantization $V$ is chosen as the in-trap volume of the gas. As a consequence of the crystalline structure of the lattice, all momentum-space quantities are periodic with the period of the reciprocal lattice $k_d=2 \pi/d$ where $d$ is the lattice spacing, {\it e.g.}, $\hat{a}(\vec{k}+\vec{K})=\hat{a}(\vec{k})$ with $\vec{K}=k_{d}(n_{x} \vec{u}_{x}+n_{y} \vec{u}_{y}+n_{z} \vec{u}_{z})$ where $\{ \vec{u}_{j} \}_{x,y,z}$ are the orthonormal vectors associated to the lattice and $\{ n_{j} \}_{x,y,z}$ are integers.\\

\subsection{Many-body correlations in a perfect Mott insulator} 

Deep in the Mott phase at large amplitude of the lattice potential, the tunnelling becomes vanishingly small and it is standard to approximate the many-body ground-state by a ``perfect" Mott insulator, {\it i.e.}, a Mott insulator with no coupling between the sites, $J=0$. The perfect Mott insulator for the Hamiltonian of Eq.~\ref{Eq:BHH} with unity occupation of the lattice site is then $| \psi \rangle_{J=0} = \Pi_{l} \ \hat{b}^{\dagger}_{l} |0\rangle$. The absence of phase coherence between the lattice sites, $\langle \psi | \hat{b}_l^{\dagger}\hat{b}_{l'} | \psi \rangle_{J=0}=\langle \hat{b}_l^{\dagger}\hat{b}_{l'} \rangle_{J=0}=\delta_{l,l'}$, has important consequences on the momentum-space properties. Under these conditions, the correlations between two momentum operators (see Eq.~\ref{Eq:MomOp}) takes the form
\begin{equation}
\langle \hat{a}^{\dagger}(\vec{k})\hat{a}(\vec{k'})\rangle_{J=0}=\frac{1}{V} \sum_{l=1}^{N_{\rm site}} \langle \hat{n}_{l} \rangle \ e^{i (\vec{k'}-\vec{k}).\vec{r}_l}. \label{Eq:akCorr}
\end{equation} 
The momentum density of a perfect Mott insulator is thus constant, $\rho(\vec{k})=\langle \hat{a}^{\dagger}(\vec{k}) \hat{a}(\vec{k}) \rangle_{J=0}=\rho(\vec{0})$. In addition, the correlations between momentum components separated by more than the inverse of the system size $L$ are vanishingly small. On the contrary, the sum in Eq.~\ref{Eq:akCorr} is non zero for $|\vec{k}-\vec{k'}|<2 \pi/L$, which implies that some correlations are present in momentum-space even if there are no position-space correlations, a situation similar to that described by the Van Cittert-Zernike theorem in Optics \cite{born1993}. Eq.~\ref{Eq:akCorr} defines the one-particle volume of coherence, {\it i.e.}, the volume over which the first-order correlation function $g^{(1)}(\vec{k},\vec{k'})=\langle \hat{a}^{\dagger}(\vec{k})\hat{a}(\vec{k'})\rangle/\sqrt{\rho(\vec{k}) \rho(\vec{k'})}$ is non-zero. This momentum-space volume of coherence reflects the distribution of atoms $\{ \langle \hat{n}_{l} \rangle \}_{l}$ in the lattice.

As pointed out in \cite{altman2004}, higher-order momentum-space correlations can reveal some properties of many-body ground-states even when the momentum density is featureless. For instance, bosonic bunching is expected in the two-body correlations $g^{(2)}(\vec{k},\vec{k'})=\langle \hat{a}^{\dagger}(\vec{k}) \hat{a}^{\dagger}(\vec{k'}) \hat{a}(\vec{k}) \hat{a}(\vec{k'}) \rangle/\rho(\vec{k}) \rho(\vec{k'})$ associated with finding one particle with a momentum $\vec{k}$ and a second one with a momentum $\vec{k'}$. In the case of the perfect Mott state, that is to say when the tunnelling is zero, the amplitude and the width of the bunching effect can be predicted accurately. Indeed, the   atom number per lattice site is then fixed. For a unity occupation of the lattice $ n_{l} =1$, the Hamiltonian of Eq.~\ref{Eq:BHH} reduces to that of non-interacting particles and is diagonal in the momentum-space basis. As a result, the many-body momentum-space correlations are those of uncorrelated bosons with a Gaussian density operator \cite{gardiner2000}. For such a Gaussian many-body ground-state, the Wick decomposition yields $g^{(2)}(\vec{k},\vec{k'})=1+|g^{(1)}(\vec{k},\vec{k'})|^2$. In particular, the amplitude of the two-body correlation at zero particle distance is twice that found for un-correlated particles, $g^{(2)}(\vec{k},\vec{k})=2$. In addition, the shape of the bunching peak provides a quantitative information about the in-trap profile. Its exact shape is set by the term $|g^{(1)}(\vec{k},\vec{k'})|^2$, whose size determines the volume of coherence. Interestingly, higher-order correlation functions are also related to the first-order correlation function $g^{(1)}(\vec{k},\vec{k'})$ when Wick theorem applies  (see \cite{dall2013} for instance). 

\subsection{Momentum density and spectral weight of a Mott insulator at finite tunnelling} 

At finite tunnelling $J>0$, a more elaborated picture than that of the perfect Mott state $| \psi \rangle_{J=0}$ must be introduced to account for the kinetic energy term in the Bose-Hubbard Hamiltonian. To do so, various many-body treatments introduce quadratic quantum fluctuations on top of the classical Gutzwiller solution for the ``perfect" Mott \cite{oosten2001, altman2002, sengupta2005}. In these approaches, the low-energy excitations of the Mott state are quasi-particles consisting of the combination of a doublon and a hole. At finite $J$, the many-body ground state contains a finite fraction of these particle/hole excitations. In addition, the mobility of the quasi-particles in the lattice restores some phase coherence. These modifications of the many-body ground-state are described by the formalism of the Green function \cite{mahan2000}. For our purposes, the Fourier transform of the Green function, {\textit{i.e.}, the one-particle spectral function $\mathcal{A}(\vec{k},\omega)$, is of special interest as it is linked to the momentum density $\rho(\vec{k})$ of the Mott insulator \cite{sengupta2005},  
$\rho(\vec{k}) \propto -\int_{-\infty}^0 d\omega \ \mathcal{A}(\vec{k},\omega)$. Replacing $ \mathcal{A}(\vec{k},\omega)$ by its expression, one finds that the  momentum density $\rho(\vec{k})$ of the Mott insulator is directly related to $Z(\vec{k})$, the spectral weight of the quasi-particles:
\begin{equation}
\rho(\vec{k})=\mathcal{N} (Z(\vec{k})-1)
\end{equation}
where $\mathcal{N}$ is a normalisation factor. The spectral weight quantifies the overlap between the many-body wave-function with one added particle (or hole) and the true excited one-particle (or one-hole) state. In the non-interacting case, $Z(\vec{k})$ is a delta function  because a state with one more particle or  hole is still an eigenstate of the system. In the correlated case, many momentum eigenstates have a non-zero overlap with the state formed by simply adding a particle or a hole. As a consequence the spectral weight acquires a finite momentum support whose shape reflects these correlations.

The momentum density $\rho(\vec{k})$ of the Mott state thus provides a direct probe of its spectral weight $Z(\vec{k})$. The spectral weight of a homogeneous Mott with unity occupation of the lattice sites can be calculated using perturbative many-body approaches \cite{sengupta2005, gerbier2005-2, freericks2009}. Deep in the Mott regime ($U/J \gg 1$), $Z(k)$ is a small-amplitude oscillating function of period $k_d=2 \pi/d$, with $d$  the lattice spacing \cite{sengupta2005, iskin2009, freericks2009}. This stems from  the presence of a small number of quasi-particles that hop between adjacent lattice sites only. On approaching the Mott transition, the spectral weight is expected to become increasingly peaked around the reciprocal lattice vectors $\vec{K}$ as an added particle (or hole) can tunnel over several lattice sites \cite{sengupta2005, iskin2009, freericks2009}. 


\section{EXPERIMENTAL RESULTS}

\subsection{Observation of the Mott transition with metastable Helium-4 atoms.} 

\begin{figure*}[ht!]
\includegraphics[width=2\columnwidth]{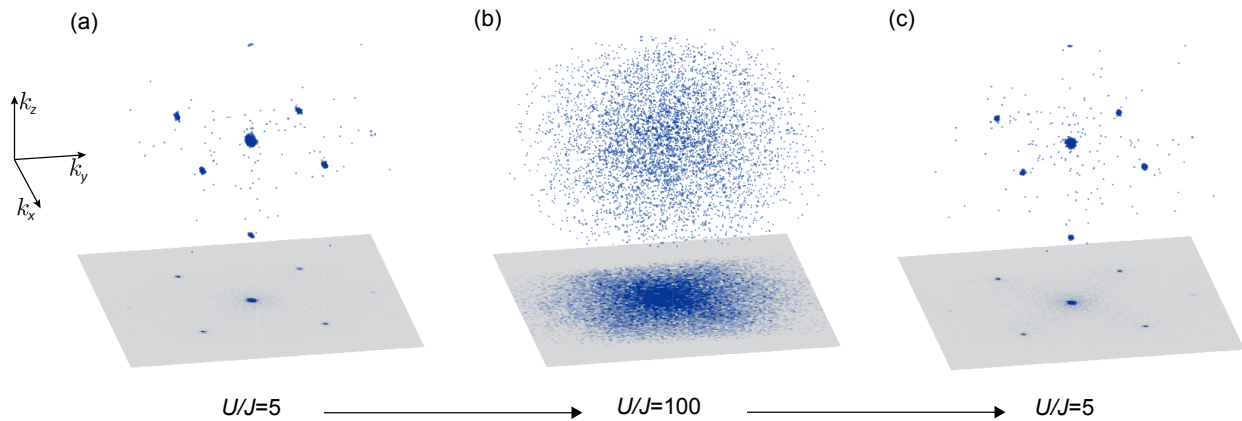}
\caption{{\bf Observation of the superfluid-to-Mott transition in momentum space with $^4$He$^*$ atoms.} Far-field three-dimensional atom distributions, along with the corresponding two-dimensional projections in the ($k_{x},k_{y}$) plane. Each blue dot is an individual atom revealed by the He$^*$ detector. {\bf (a)} Superfluid regime at $U/J=5$ , {\bf (b)}  Mott regime $U/J=100$, {\bf (c)} superfluid regime $U/J=5$ restored when sweeping back from $U/J=100$ to $U/J=5$. With this set of data,  the visibility of the interference pattern \cite{gerbier2005} goes from $\mathcal{V}=0.96$ in figure {\bf (a)}  to $\mathcal{V}=0.92$   in figure {\bf (c)}. This demonstrates that the loading ramps are adiabatic with a good approximation.}
\label{Fig1}
\end{figure*}

The experimental sequence starts with the production of a  Bose-Einstein condensate (BEC) of metastable Helium-4 atoms ($^4$He$^*$) in a crossed optical trap as described in \cite{bouton2015}. The BEC is then loaded into the lowest energy band of a 3D cubic optical lattice with a spacing $d=775~$nm and an amplitude $V_0=s E_r$, with $E_r=h^2/8 m d^2$ the recoil energy \cite{cayla2018}. The lattice amplitude $s$ is increased linearly at a rate $0.3~$ms$^{-1}$ while, simultaneously, the intensity of the optical trap is switched off with a linear ramp of duration $20~$ms. The residual harmonic potential due to the Gaussian intensity profiles of the orthogonal lattice beams is isotropic with a trapping frequency $140 \sqrt{s}$ Hz. We calibrate the amplitude $s$ by performing amplitude modulation spectroscopy and we numerically calculate $J$ and $U$ from the value $s$. The rms uncertainty on $U/J$ is estimated to be 5\%. 

In the experiment we measure time-of-flight single-atom-resolved distributions in far-field, from which we can calculate the 3D momentum density as well as higher-order correlation between individual atoms. We probe the gas after a time-of-flight of $t_{\rm tof}=297$~ms with the time- and space-resolved} He$^*$ detector described in~\cite{nogrette2015}. The  He$^*$ detector reveals the 3D positions $\vec{r}_{\rm tof}$ of individual $^4$He$^*$ atoms in the center-of-mass reference frame with an excellent resolution (see Annex B). Importantly, $t_{\rm tof}$ is larger than the time required to enter the far-field regime of propagation $t_{\rm FF}=m L^2/2\hbar \sim 30$ ms for our experimental parameters where $L \sim 40 d$ is the in-trap total size of the gas. In addition, interaction can be neglected during the expansion as the TOF dynamics is driven by the large trapping frequency ($\gtrsim100~$kHz) of the individual lattice sites \cite{gerbier2008}. Indeed, in a previous work \cite{cayla2018}, we validated quantitatively the assumption of ballistic expansion from the lattice, {\it i.e.}, $\hbar \vec{k}=m \vec{r}_{\rm tof}/t_{\rm tof}$ where $\vec{k}$ is the momentum of the atom in the lattice, $\vec{r}_{\rm tof}$ is the measured atom position after a time-of-flight $t_{\rm tof}$. With the ballistic assumption, the far-field TOF density $\rho_{\infty}(\vec{r}_{\rm tof},t_{\rm tof})$ yields the in-trap  momentum density $\rho(\vec{k})$, 
\begin{equation}
\rho(\vec{k})= \frac{1}{|\tilde{\omega}(\vec{k})|^2} \left ( \frac{\hbar t_{\rm tof}}{m} \right )^3  \rho_{\infty}(\vec{r}_{\rm tof},t_{\rm tof}) \label{Eq:MomDen}
\end{equation}
where $\tilde{\omega}(\vec{k})$ is the Fourier transform of the Wannier function in each lattice site. We determine $\tilde{\omega}(\vec{k})$ numerically in 3D and rescale the measured density $\rho_{\infty}(\vec{r}_{\rm tof},t_{\rm tof})$ according to Eq.~\ref{Eq:MomDen}. Doing so, $\rho(\vec{k})$ can be directly compared to the momentum density of the {\it discrete} Bose-Hubbard Hamiltonian introduced in the theory section (see Eq.~\ref{Eq:BHH}). 

In Fig.~\ref{Fig1} we plot examples of 3D TOF distributions across the Mott transition. The Mott transition has been investigated with various atomic species \cite{bloch2012} and here we report its observation with metastable Helium-4 atoms which allows us to obtain 3D single-atom-resolved distributions. Each blue dot in Fig.~\ref{Fig1} corresponds to one detected atom. In Fig.~\ref{Fig1}a, the distribution displays sharp diffraction peaks located at $k_d=2\pi/d$, a manifestation of the long-range coherence of the superfluid state at $U/J=5$. Deep in the Mott insulating regime ($U/J=100$), the phase coherence is lost (see Fig.~\ref{Fig1}b). By decreasing $U/J$ from the Mott state back to the superfluid regime, the long-range coherence is restored, as illustrated by the high visibility of the diffraction peaks in Fig.~\ref{Fig1}c. In the following, we focus on the Mott insulator regime using values of the ratio $U/J$ larger than the critical value for the Mott transition $(U/J)_{\rm c}=29.3$ \cite{capogrosso2007}

\subsection{Many-body momentum-space correlations deep in the Mott insulating regime} 

\begin{figure}[h!]
\includegraphics[width=\columnwidth]{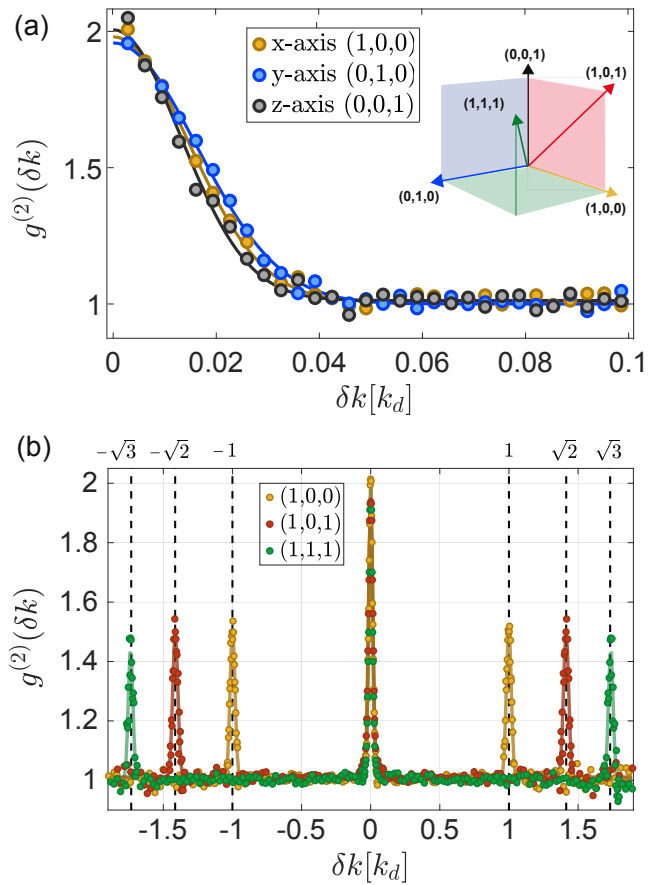}
\caption{{\bf Two-body momentum correlations in bosonic Mott insulators at $U/J=100$ with $N=15~000$}. \textbf{(a)} Plot of 1D cuts through the two-particle momentum correlation $g^{(2)}(\delta k)$ as a function of $\delta k=\delta \vec{k}.\vec{u}$ along the axis defined by $\vec{u}= n_x \vec{u}_x + n_y \vec{u}_y + n_z \vec{u}_z$. The values of the integers $(n_{x},n_{y},n_{z})$ are indicated next to the symbols of the data. The elementary volume we use to code the positions of individual atoms has a longitudinal length $\Delta k_{||}=0.003~ k_d$ along the axis where correlations are plotted and a transverse length $\Delta k_{\perp}=0.015~k_d$. The 1D cuts are plotted along the three axis of the cubic lattice, illustrating the spherical symmetry of the experimental configuration. \textbf{(b)} 1D cuts $g^{(2)}(\delta k)$ plotted on a broader scale through the $g^{(2)}(\delta \vec{k})$ along different axis of the reciprocal lattice.}
\label{Fig2}
\end{figure}

We first discuss the single-atom-resolved distributions measured deep in the Mott insulator regime. As expected, we find that the momentum density $\rho(k)$ (as defined in Eq.~\ref{Eq:MomDen}) is almost constant in this regime. To characterize the fluctuations of the momentum-space population, we introduce the two-particle correlation function defined as (see Annex C for details),
 \begin{equation}
 g^{(2)}(\delta \vec{k})=\frac{\int d\vec{k} \ \langle \hat{a}^{\dagger}(\vec{k}) \hat{a}^{\dagger}(\vec{k}+\delta \vec{k}) \hat{a}(\vec{k}+\delta \vec{k})\hat{a}(\vec{k})  \rangle}{\int d\vec{k}\ \rho(\vec{k}) \rho(\vec{k}+\delta \vec{k})}
 \label{Eq:correlation}
 \end{equation}
 where $\hat{a}(\vec{k})$ is the annihilation operator associated to finding a particle at position $\vec{k}$ with the He$^*$ detector. To facilitate the presentation and the discussion of the two-body correlation function, we plot $g^{(2)}(\delta \vec{k})$ along some specific directions $\vec{u}$ that can be chosen at will.  In Fig.~\ref{Fig2}, we plot $g^{(2)}(\delta \vec{k})$  in a Mott insulator with $U/J=100$ along $ \vec{u}\propto n_x \vec{u}_x + n_y \vec{u}_y + n_z \vec{u}_z$ where $\{ \vec{u}_{j} \}_{x,y,z}$ are the orthonormal vectors associated to the lattice and $n_x$, $n_y$ and $n_z$ are equal to 0 or 1. The two-body correlation function is plotted as a function of the difference $\delta k$ of momentum between two atoms found along the $\vec{u}$ axis (see Annex C). We find a well contrasted bunching whose amplitude $g^{(2)}(0)=1.98(3)$ is close to two, as expected for Gaussian statistics. The corresponding large fluctuations in the momentum-space are in contrast with the reduced atom number fluctuations in the lattice sites, as monitored with in-situ quantum gas microscopes in a similar regime of Mott insulator \cite{sherson2010}. Note that there has been only a few observations of $g^{(2)}(0)\simeq 2$ with massive particles \cite{ottl2005, dall2013, lopes2014} and that none of them was obtained in a 3D isotropic situation. As illustrated in Fig.~\ref{Fig2}(b), the two-body correlation $g^{(2)}(\delta \vec{k})$ is 3D-periodic. We find periods equal to $k_d$, $\sqrt{2}k_{d}$ and $\sqrt{3}k_{d}$ along the different of the 3D reciprocal lattice associated with the crystalline structure of the ground-state. As we shall explain below, the lower amplitude of the bunching peaks at $\delta \vec{k}\neq \vec{0}$ is due to small imperfections of the He$^*$ detector.
 
We now turn to investigating the shape and width of the two-body correlation function. In the vicinity of $\delta \vec{k}=\vec{0}$, we find that $g^{(2)}(\delta \vec{k})$ is well fitted by an isotropic 3D Gaussian function. This isotropy corresponds to the isotropy of the optical trap created by the lattice beams. The Gaussian-like shape probably results from the spherical distribution of atoms in the lattice and from the absence of sharp edges in the in-trap density. These observations about the isotropy and the shape of the measured two-body correlations lead us to define the two-particle correlation length $l_{c}$ according to $g^{(2)}(\delta \vec{k})=1+\eta \exp(-2 \delta \vec{k}^2/l_{c}^2)$  at $|\delta \vec{k} | \ll k_d$, where $\eta$ corresponds to the amplitude of the bunching. We find a correlation length $l_{c}=0.027(6)k_{d}$ for the data of Fig.~\ref{Fig2}(a).

\begin{figure}[ht!]
\includegraphics[width=\columnwidth]{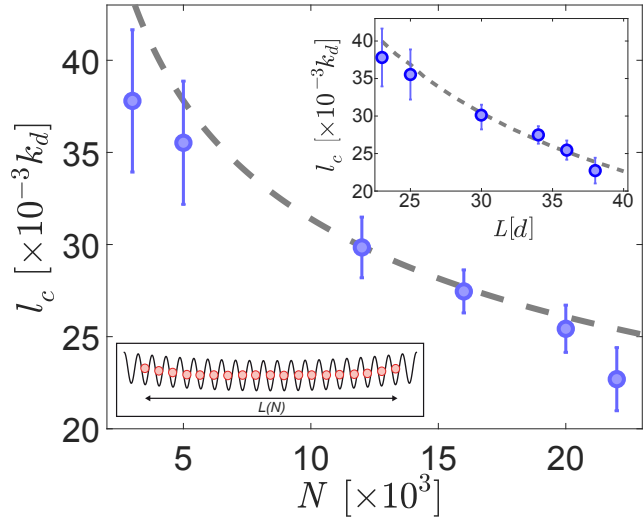}
\caption{{\bf Two-body correlation length.} Correlation length $l_{c}$ plotted as a function of the total atom number $N$ in the Mott state. The dashed line is the theoretical prediction from 3D numerical calculations without adjustable parameters. Inset: $l_{c}$ as a function of the calculated in-trap total size $L$. The dashed line is the same numerical calculation as that of the main panel.}
\label{Fig3}
\end{figure}

As discussed in the theory section, the two-body correlations of a perfect Mott state are known accurately as one can compute $g^{(2)}(\delta \vec{k})=1+|g^{(1)}(\delta \vec{k})|^2$ from the sum appearing in Eq.~\ref{Eq:akCorr}. This requires the knowledge of the spatial distribution of atoms in the 3D lattice, which we obtain using the Gutzwiller ansatz for the experimental parameters \cite{jaksch1998}. To investigate in the experiment how the shape of $g^{(2)}(\delta \vec{k})$ is modified when the spatial distribution of the atoms in the 3D lattice varies, we use the low compressibility of the gas: the size $L$ of a trapped Mott insulator increases with the atom number $N$ as a result of the strong on-site interactions. We have varied the atom number from $N=3.0(6)\times 10^3$ to $N=22(4) \times 10^3$ and measured the different correlation lengths $l_{c}$. Over this range, we note that the lattice filling at the trap center varies from 1 to 2, but numerical calculations of $g^{(2)}$ show that the presence of a few doubly-occupied sites at the trap centre hardly affects the value of $l_{c}$. In Fig.~\ref{Fig3} we plot the measured $l_{c}$ as a function of the atom number $N$. The two-body correlation length $l_{c}$ is found to decrease with $N$, as expected from the fact that $L$ increases with $N$. We also plot $l_{c}$ as a function of the size $L$ obtained from the Gutzwiller ansatz in the inset of Fig.~\ref{Fig3} (see Annex C for details). Moreover, the measured values for $l_{c}$ are in perfect agreement with numerical calculations without adjustable parameters, {\it i.e.}, with the expectation for the atom distribution of a Mott state and for a Gaussian density operator. Note that in a thermal gas of non-interacting bosons, the size $L$, and in turn $l_{c}$, are independent of $N$. The observed variation of $l_{c}$ with $N$ thus highlights the difference between probing the Mott state and probing an ideal Bose gas \cite{dall2013}. As a side remark, we note that an analogy with the Hanbury-Brown and Twiss effect \cite{hanbury1956, folling2005} can also be used to describe the observed two-body correlations (see Annex A for details).

Proceeding similarly with a Gaussian fit, we measure the two-body correlation length $l_{c}^{\delta k\neq 0}=0.033(9) k_{d}$ for the peaks of Fig.~\ref{Fig2}(b) observed at $\delta \vec{k} \neq \vec{0}$. This value is slightly larger than $l_{c}$ (measured at $\delta \vec{k}=\vec{0}$), by a few times $k_{a}/1000$. This tiny discrepancy may be explained by small imperfections of the He$^*$ detector in measuring particle distances comparable to its radius (see for instance \cite{jagutzki2002, hong2016} and Annex C). While the measured correlation lengths and amplitudes may be affected by the response function of the detector, the 3D integral of the bunching peak, which quantifies the probability for two atoms to bunch, should be a physical quantity conserved through the detection. In a crystalline ground-state as a Mott insulator, atoms are expected to have the same probability to bunch modulo a vector $\vec{K}$ of the reciprocal lattice. This implies that the 3D integrals of the bunching peaks should be equal on the reciprocal lattice. Importantly, we find that this property is verified in the experiment (see Annex C). This implies that the observed amplitude $g^{(2)}(\vec{K}\neq\vec{0})<2$ results from the measured correlation length $l_{c}^{\delta k\neq 0}$ larger than $l_{c}$. 
 
\begin{figure}[ht!]
\includegraphics[width=\columnwidth]{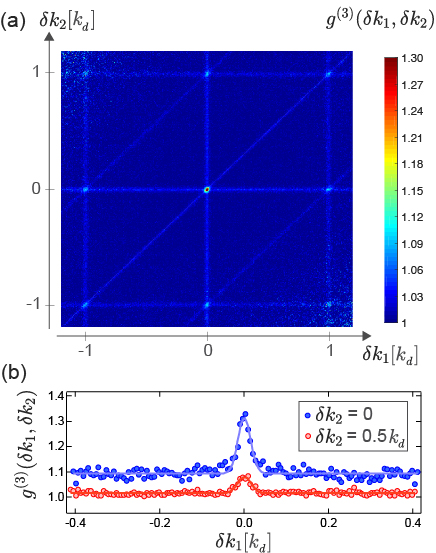}
\caption{{\bf Three-body momentum correlations in bosonic Mott insulators.} {\bf (a)} Plot of $g^{(3)}(\delta k_{1}, \delta k_{2})$ as a function of the algebraic distances $\delta k_{1}$ and $\delta k_{2}$ between atoms along the lattice axis (see text). The background amplitude associated to un-correlated atoms has an amplitude $\sim 1$. The correlations along the lines defined by $\delta k_{1}=0$, $\delta k_{2}=0$ and $\delta k_{1} =\delta k_{2}$ correspond to the two-body bunching $g^{(2)}(0)$ associated with finding two atoms close-by. The intrinsic three-body correlations (located at $\delta k_{1}=0[k_d],\delta k_{2}=0[k_d]$) have an even higher amplitude. Here we used $\Delta k_{||}=0.008~ k_d$ and $\Delta k_{\perp}=0.1~k_d$. {\bf (b)} Profiles of $g^{(3)}(\delta k_{1}, \delta k_{2})$ at $ \delta k_{2}=0$ and $ \delta k_{2}=0.5 k_{d}$. We find $g^{(3)}(0,0)-1=0.32(2)$ and $g^{(3)}(0,0.5 k_{d})=0.065(10)$, {\it i.e.} that the ratio between the amplitudes of three-body and two-body correlations is compatible with the expected value for Gaussian statistics.}
\label{Fig4}
\end{figure}

From the measured 3D atom distributions, we can extract higher-order correlations. To illustrate this possibility, we have measured the three-body correlations. We observe a well contrasted and periodic bunching in the three-body correlations $g^{(3)}(\vec{k}, \vec{k'}, \vec{k''})$ whose numerator writes $\langle \hat{a}^{\dagger}(\vec{k}) \hat{a}^{\dagger}(\vec{k'}) \hat{a}^{\dagger}(\vec{k''}) \hat{a}(\vec{k})  \hat{a}(\vec{k'}) \hat{a}(\vec{k''}) \rangle$. In Fig.~\ref{Fig4}(a) we plot $g^{(3)}(\delta k_{1},  \delta k_{2})$ as a function of $\delta k_{1}$ and $\delta k_{2}$  where $\delta k_{1}=(\vec{k}-\vec{k'}).\vec{u}_{x}$ and $\delta k_{2}=(\vec{k}-\vec{k''}).\vec{u}_{x}$ are the differences of momentum between two different pairs of atoms along the lattice axis $\vec{u}_{x}$. The correlations that are intrinsically three-body are those for which three atoms are in the same coherence volume and thus they are located at positions $(\delta k_{1}=0[k_d],\delta k_{2}=0[k_d])$ where $[k_d]$ stands for modulo $k_d$. We also observe lines with a correlation above the background along the axis defined by $\delta k_{1}=0$, by $\delta k_{2}=0$ and by $\delta k_{1}=\delta k_{2}$. Along each of these lines, two of the three atoms are close to each other and the correlation amplitude signals two-body bunching. No correlation is observed on the anti-diagonal defined by $\delta k_{1}=-\delta k_{2}$. This is expected since it corresponds to well separated, thus un-correlated, momentum components. Finally, the ratio of the amplitudes of the intrinsic three-body correlations $g^{(3)}(0,0)$ and the two-body ones $g^{(3)}(0,\delta k_{2}\neq0)$ is found to be 2.95(40), a value compatible with the expected one $3!/2!=3$ for Gaussian statistics \cite{dall2013}. Note that this ratio was obtained for transverse integrations ($\Delta k_{\perp}=0.1 k_{d}$) larger than those used in Fig.~\ref{Fig2} in order to increase the statistics of the three-body correlations. As a consequence, the measured absolute amplitudes are smaller than the ones expected in the absence of integration (Fig.~\ref{Fig4}(b)).

\subsection{Quasi-particle Spectral weight in a Mott insulator} 

\begin{figure*}[ht!]
\includegraphics[width=2\columnwidth]{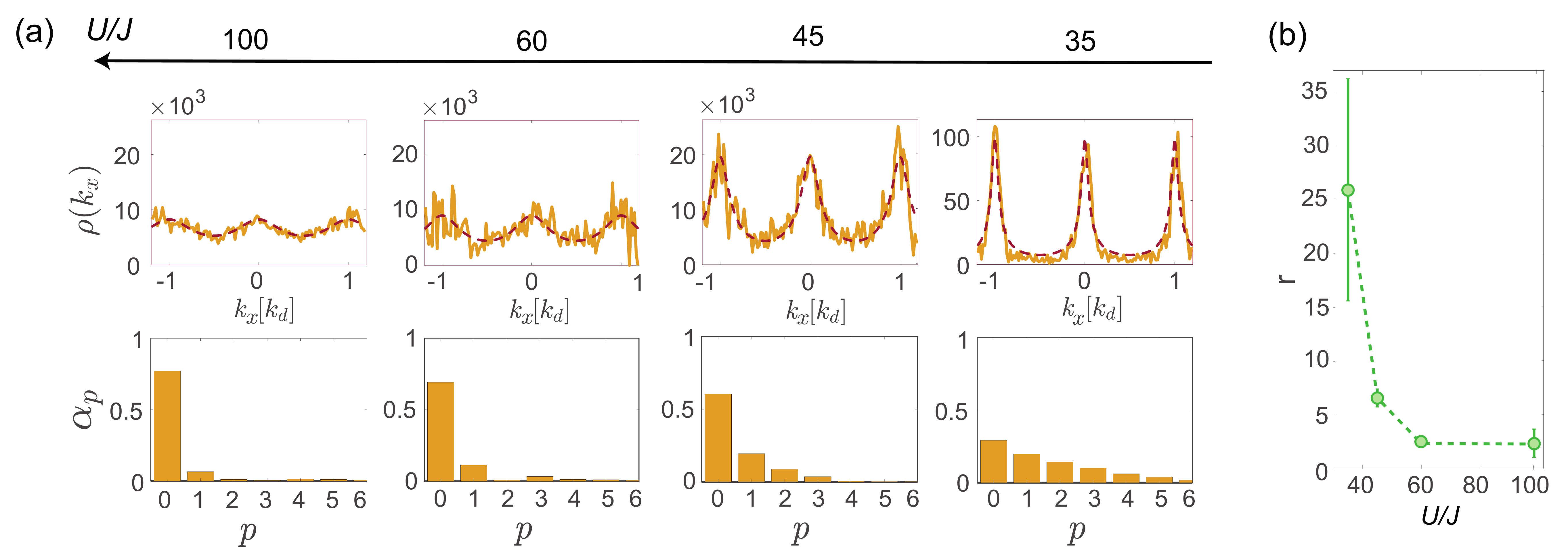}
\caption{{\bf Spectral weight of the quasi-particles in the Mott state.} {\bf (a)} Top row: Dimensionless momentum densities $\rho(k_{x})\propto Z(k_{x})-1$ in the Mott regime for a ratio $U/J=100$, $60$, $45$, $35$. The dashed line is the momentum density expected from perturbation many-body theories accounting for the presence of particle-hole excitations, with the value of $U/J$ taken as a fitting parameter \cite{freericks2009}. Bottom row: histograms of the amplitudes $\alpha_{p}$ corresponding to the momentum densities displayed on top. {\bf (b)} Ratio $r=[U/J-(U/J)_{c}]/[u_{\rm fit}-(U/J)_{c}]$ plotted as a function of $U/J$. The ratio $r$ quantifies the discrepancy between $U/J$ set in the experiment and $u_{\rm fit}=U/J$ obtained from fitting the experimental data with perturbative theories.}
\label{Fig5}
\end{figure*}

When $U/J$ is decreased, a structure slowly appears in the momentum density. Measuring momentum 3D densities in far-field and without the line-of-sight integration inherent to absorption imaging allows us to investigate this evolution quantitatively. In Fig.~\ref{Fig5}(a) we plot $\rho(k_{x})$ in the Mott regime along the lattice axis $Ox$ and for a varying ratio $U/J$. All these data sets have been taken with a fixed atom number $N=3.0(6)\times 10^3$ to ensure unity occupation of the lattice at the trap center. At large $U/J$ we recover the periodic oscillations observed previously \cite{gerbier2005, spielman2007}. For decreasing $U/J$, we observe that $\rho(k_{x})$, and thus the spectral weight $Z(k_{x})\propto \rho(k_{x}) +1$, becomes increasingly peaked around $k_{x}=j k_d$ with $j$ integer. To provide a physical insight into the narrowing of $Z(k_{x})$, we expand $\rho(k_{x})=\sum_{p} \alpha_{p} \ {\rm e}^{i 2 \pi p \ k_{x}/k_d}$ in Fourier components along the lattice axis, finding
\begin{equation}
\alpha_{p}=  \sum_{j} \langle \hat{b}_{j}^\dagger \hat{b}_{j+p} \rangle.
\end{equation}
The Fourier amplitudes $\alpha_{p}$ quantify the average phase coherence between lattice sites distant by $p$ sites. The evolution of the amplitudes $\alpha_{p}$ as $U/J$ decreases highlights the continuous change from a perfect Mott state (for which $\alpha_{p}=0$ for $p\neq0$) to a Mott state with a finite spatial coherence ($\alpha_{p} \neq 0$ for $p>1$). Close to the Mott transition, the mobility of the quasi-particles extends over several lattice sites, say distant by $p$ sites, and results in $\alpha_{p} \neq 0$. The narrowing of the spectral weight $Z(k_{x})$ thus reflects the coherent tunnelling of the quasi-particles over several lattice sites. In Fig.~\ref{Fig5}(a) (bottom row) we plot the Fourier amplitudes $\alpha_{p}$ extracted from the measured profiles $\rho(k_{x})$. A mobility of the quasi-particles over up to $p \sim 6$ lattice sites is observed at $U/J=35$, a distance corresponding to about a quarter of the size of the trapped gas.

The spectral weight of a homogeneous Mott with unity occupation of the lattice sites was calculated using perturbative many-body approaches \cite{sengupta2005, gerbier2005-2, freericks2009}. It provides a quantitative prediction of the role of particle-hole excitation on the building of long-range coherence which we confront with the experiment. To do so, we fit the measured densities $\rho(k_{x})$ with the equation Eq.~(95) of \cite{freericks2009} using the numerical parameters associated with a 3D lattice and letting the ratio $U/J=u_{\rm fit}$ as a fit parameter ({\it i.e.} the parameter $x$ of \cite{freericks2009}). The observed modification in the shape of $Z(k_{x})$ as the system approaches the superfluid regime is consistent with these theoretical predictions (see Fig.~\ref{Fig5}(a)). As the predicted width and amplitude of the modulations are connected through the dispersion of the lattice, this agreement suggests that the physical picture drawn by perturbative theories is valid. 

However, a fully quantitative comparison reveals that $u_{\rm fit}$ is systematically smaller than the ratio $U/J$ calibrated in the experiment. To quantify the discrepancy between theory and experiment, we define a fractional deviation as the ratio $r=[U/J-(U/J)_{c}]/[u_{\rm fit}-(U/J)_{c}]$ and we plot $r$ as a function $U/J$ in Fig.~\ref{Fig5}(b). We find that $r$ is large close to the critical ratio $(U/J)_{c}$ and that it decreases as the ratio $U/J$ increases, becoming closer to unity. Our understanding of this observation goes at follows. Close to the Mott transition, the trapped system realised in the experiment is composed of a Mott insulator phase surrounded by a condensate. The later obviously increases the coherence of the trapped system with respect to that predicted by the theory. As $U/J$ increases, the Mott region with unity filling extends over a larger volume in the trap and the outer-shell becomes normal. As a result, $r$ should become closer to unity at increasing $U/J$. However, we do not observe $r =1$ at large $U/J$, but $r\simeq2$, a difference that may be related to the increasing difficulty to adiabatically load the atoms in the lattice at increasing $U/J$. We note that the observed trend for the ratio $r$ differs from that found in a previously reported work with a similar quantity, the visibility \cite{gerbier2005-2}. This is because the theoretical model we use here, which captures the position of the Mott transition predicted by QMC calculations, is different from that used in \cite{gerbier2005-2}.

\section{CONCLUSION}

We have reported on the measurement of 3D atom distributions in momentum-space for a Mott insulator. It has allowed us to determine quantitatively the build up of long range coherence when approaching the transition towards a superfluid. Deep in the Mott phase, the extracted two- and three-body correlation functions match that of indistinguishable and uncorrelated bosons uniformly distributed in the lattice sites. It shows that the measured correlations are that of a many-body state whose density operator is Gaussian in momentum-space. The results of this work also demonstrate the outstanding capability of our approach to identify Gaussian quantum states. This naturally paves the way to the future investigation of 3D many-body states with many-body correlations deviating from those associated with Gaussian statistics. As illustrated with 1D Bose gases \cite{schweigler2017}, these deviations provide genuine information about many-body correlations and have therefore the potential to unveil complex quantum phenomena.    


\begin{acknowledgments}
We acknowledge fruitful discussions with D. Boiron, F. Gerbier, T. Roscilde and all the members of the Quantum Gases group at Institut d'Optique. We acknowledge financial support from the LabEx PALM (Grant number ANR-10-LABX-0039), the R\'egion Ile-de-France in the framework of the DIM SIRTEQ and the Agence Nationale pour la Recherche (Grant number ANR-17-CE30-0020-01). D.C. acknowledges support from the Institut Universitaire de France.
\end{acknowledgments}

\begin{center}{\bf APPENDIX A: OPTICAL ANALOGY OF THE MOMENTUM-SPACE BUNCHING} \end{center}

The description of the momentum-space two-body correlations we investigate in this work could be rephrased in an optical analogy. Indeed, the momentum-space properties were obtained in a time-of-flight (TOF) experiment where the matter-wave evolves freely in space similarly to the propagation of light. When the expansion of the gas is ballistic and long enough for the paraxial approximation to be valid, the atoms are detected in a regime identical to that of the detection of photons in the far-field. The momentum-space bunching phenomenon can thus be understood from a direct analogy with the Hanbury-Brown and Twiss (HBT) effect reported with incoherent sources of light \cite{hanbury1956} or matter \cite{schellekens2005, folling2005}. 

It is also interesting to recall that the properties of the HBT effect in Optics may be derived from a classical model of light with Gaussian statistics \cite{goodman1985}. This requires to assume that a large number of mutually-incoherent emitters form the source of light, a condition under which the Central Limit Theorem applies and the field is a Gaussian random process. This assumption is a classical analog of the hypothesis of a Gaussian ground-state we use here, and it highlights the fact that in both situations HBT-type of correlations are connected with Gaussian statistics.
 
\begin{center}{\bf APPENDIX B: ATOM NUMBER CALIBRATION AND HE$^*$ DETECTOR} \end{center}

\begin{center}
\textit{Calibration of the  atom number in the lattice}
\end{center}

The data presented in this work comes from various sets of the experiment: 
\begin{itemize}
\item a series of sets taken with $N=3.0(6) \times 10^3$ atoms in the lattice and for $U/J=35$, $45$, $60$ and $100$. This atom number ensures a unity occupation of the lattice sites at the trap center and for all the lattice amplitudes. These sets were used for Fig.~\ref{Fig1}, Fig.~\ref{Fig4} and for the points at $L=23$ and $L=25$ in Fig.~\ref{Fig3}.
\item a set of data at $U/J=100$ with $N=15\times 10^3$ atoms, corresponding to a filling at the trap center equal to 2. This larger atom number increases the signal-to-noise ration to measure the two-body correlation length. We checked by performing numerical simulations that the bunching peaks are quite insensitive to the presence of a few doubly-occupied sites and that $l_{c}$ is dominated by the width $L$ of the trapped gas. 
Taking into account atom number fluctuations in the experiment, the overall data have been divided in 4 sets of data with $N=12(3)\times 10^3$, $N=16(3)\times 10^3$, $N=20(4)\times 10^3$  and $N=22 (4)\times 10^3$  atoms . 
\end{itemize}

To calibrate the atom number,  Mott insulators at $U/J=100$ were probed by absorption imaging  after $1.5$ ms time of flight (TOF). At this lattice amplitude, the momentum distribution is diluted and absorption images can be fitted by a 2D gaussian function from which one can extract the number of atoms.
\newline

\begin{center}
\textit{Detection with the He$^*$ detector}
\end{center}

To access single-atom-resolved momentum distributions, the atoms are detected on the  He$^*$ detector located $43$ cm below the center of the science chamber, corresponding to a time of flight  of $297$ ms. To avoid the effect of spurious magnetic fields that would disturb the time of flight  distribution, a fraction of the atoms, initially spin-polarized into the $2^3S_1$ $m_J=1$ state, are transferred with a radio-frequency (RF) resonant pulse to the non-magnetic $m_J=0$ state at the beginning of the TOF. The atoms remaining in a state with a non-zero magnetic moment are expelled from the detection area by the use of a magnetic gradient. By adjusting the duration of the RF pulse, the fraction of atoms transferred to the non-magnetic state, in which atoms are detected, can be controlled. To make sure that there is no saturation effect of the He$^*$ detector, we choose to detect about 5\%  of the atom number per shot. Between 200 and 2~000 shots were then taken for each specific measurement.

\begin{center}{\bf APPENDIX C: CORRELATION FUNCTIONS EXTRACTED FROM THE MEASURED ATOM DISTRIBUTIONS}\end{center}

The measurement of  two-particle momentum correlations quantifies the conditional probability for an atom in a given experimental realization to have a momentum $\vec{k}_1$ provided one atom is detected with a momentum $\vec{k}_2$. 
\begin{equation}
 G^{(2)}(\vec{k}_1,\vec{k}_2)=<\hat{a}^{\dagger}(\vec{k}_1) \hat{a}^{\dagger}(\vec{k}_2) \hat{a}(\vec{k}_2) \hat{a}(\vec{k}_1)>
\end{equation}
In our experiment, the measurement of the full 3D momentum distribution with single atom sensitivity allows to compute the 3D correlation functions. However, as the representation of the full distribution is intricate, we calculate two-body correlations along some specific directions $\vec{k}_1-\vec{k}_2\propto \vec{u}$. These directions can be chosen at will and in the present work they are oriented along the reciprocal lattice vector $ \vec{u}\propto n_x \vec{u}_x + n_y \vec{u}_y + n_z \vec{u}_z$ 
with $n_x$,$n_y$ and $n_y$ integers and $\vec{u}_x$, $\vec{u}_y$ and $\vec{u}_z$ are the lattice axis. The translation invariance of the state we probe implies that $ G^{(2)}$ depends only on the momentum separation between two particles,  \textit{e.g}.  $ G^{(2)}(\vec{k}_{1},\vec{k}_{2})= G^{(2)}(\delta k~ \vec{u})$ with $\vec{k}_1-\vec{k}_2=\delta k ~ \vec{u}$. We consequently calculate $G^{(2)}$ along the lattice direction by integrating over the position of one of the two atoms: 
\begin{equation}
 G^{(2)}(\delta \vec{k})=\int_{\vec{k}} <\hat{a}^{\dagger}(\vec{k}) \hat{a}^{\dagger}(\vec{k}+\delta \vec{k}) \hat{a}(\vec{k}+\delta \vec{k}) \hat{a}(\vec{k})> ~ d\vec{k}.
\end{equation} 
Here the integral refers to the summation over all the atoms  present in one shot of the experiment. To increase the signal-to-noise ratio, we also perform a transverse integration: 
\begin{equation}
G^{(2)}(\delta k\vec{u})=\int_{|\vec{k}_{\perp}|<\Delta k_{\perp}}G^{(2)}(\delta k\vec{u}+\vec{k}_{\perp}) d\vec{k}_{\perp}
\end{equation} 

\begin{figure}[ht!]
\includegraphics[width=\columnwidth ]{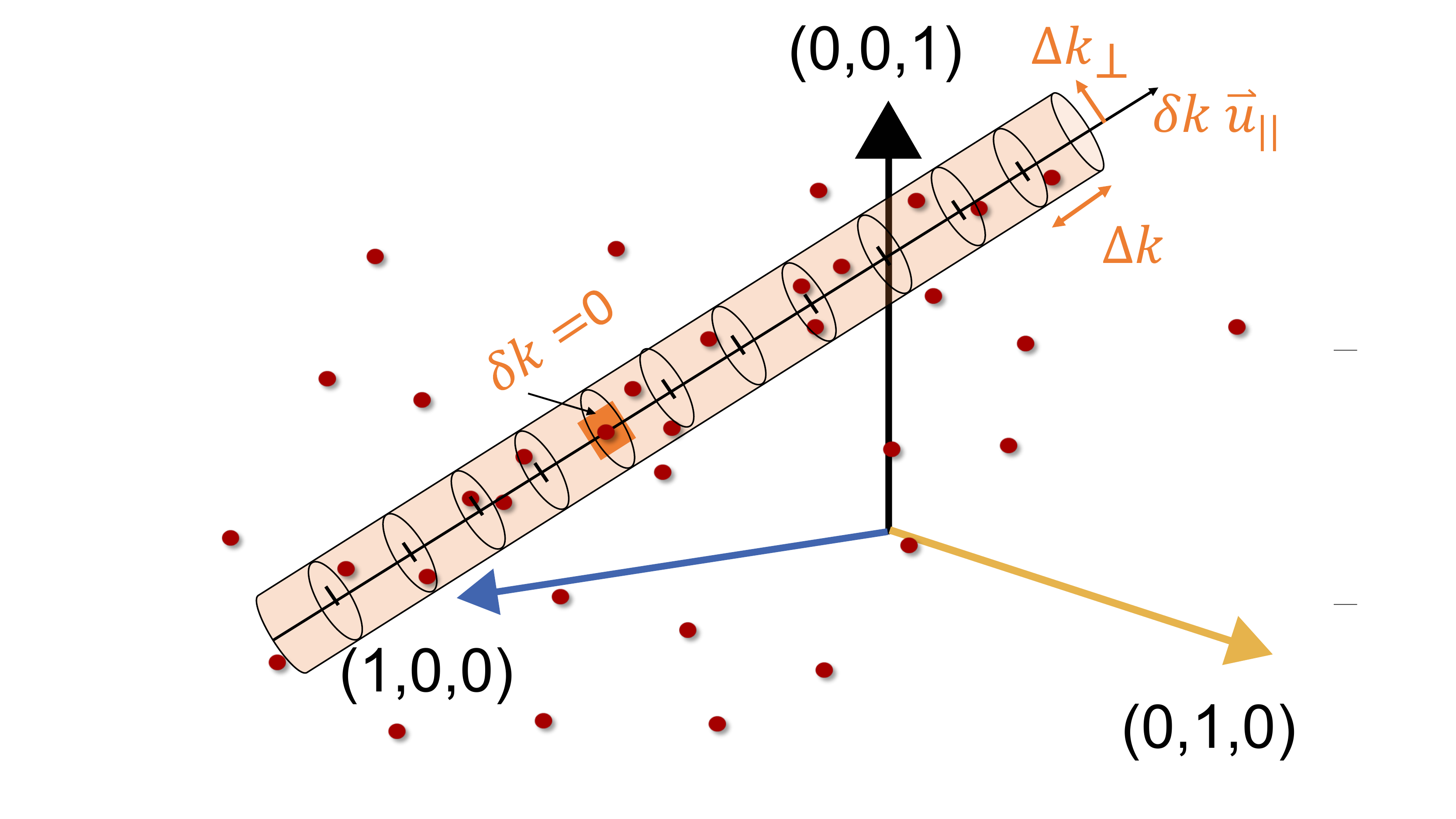}
\caption{\textbf{Method to calculate $G^{(2)}$.} One atom from a given shot is chosen (highlighted in orange). A tube of radius $\Delta k_{\perp}$ oriented along $\vec{u}$ and centered on the considered atom is defined. The distances $\delta k$ between the atom and the other ones contained in the tube are extracted. The operation is repeated for all the atoms of the shot and the results are saved in a histogram with a longitudinal step $\Delta k$. }
\label{Fig:Sup1}
\end{figure}

\begin{figure}[ht!]
\includegraphics[width=\columnwidth]{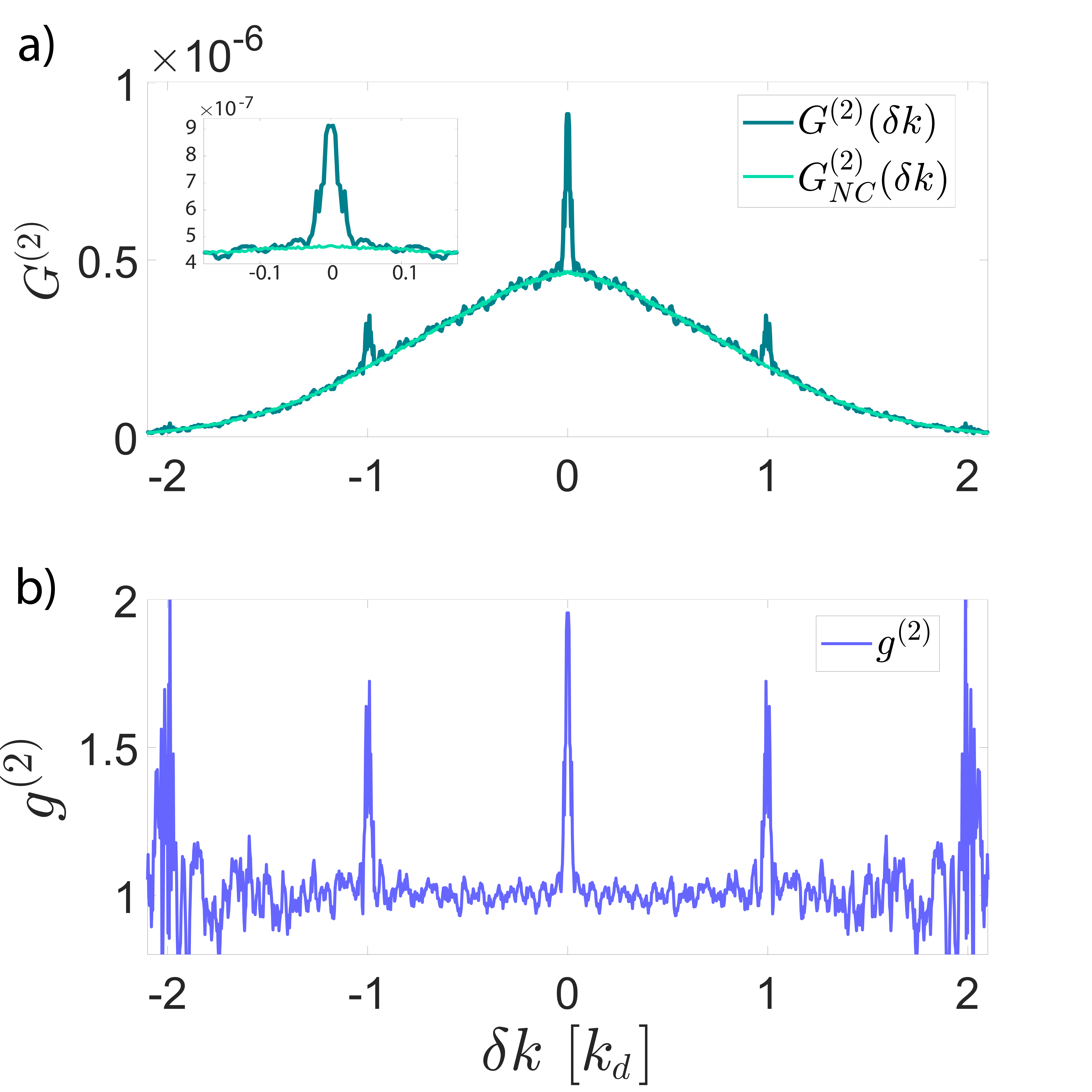}
\caption{\textbf{Correlation functions.}  Value of the different correlation functions  along the x-axis for the data displayed in Fig.~\ref{Fig2}. \textbf{a)}~ $G^{(2)}(\delta k)$ and $G_{\text{NC}}^{(2)}(\delta k)$. \textbf{b)}~  $g^{(2)}(\delta k)$.}
\label{Fig:Sup4}
\end{figure}

The procedure to calculate $G^{(2)}(\delta k)$ is depicted in Fig.~\ref{Fig:Sup1}. For each atom, we define a tube of radius $\Delta k_{\perp}$ oriented along $\vec{u}$ from which the histogram of the distances from this atom to the ones contained in the tube is recorded. The histograms corresponding to the different atoms of the shots are then summed and $G^{(2)}$ averaged over many realizations of the experiment. A plot of $G^{(2)}$ measured along the x-axis and corresponding to the data displayed in Fig.~\ref{Fig2} is given in Fig.~\ref{Fig:Sup4}a. Three bunching peaks  are visible at $\delta k=0$ and $\delta k=\pm k_d$ on top of a broad background which can be identified as resulting from the Fourier transform of the Wannier function. This background is equal to the autocorrelation of the momentum density of the Mott insulator. It corresponds to the value of the $G^{(2)}$ function in the absence of correlation between atoms and later referred as $G_{\text{NC}}^{(2)}$. The ratio $G^{(2)}$ divided by $G_{\text{NC}}^{(2)}$ yields the normalized two-body correlation function $g^{(2)}$ used in the main text. By definition, $g^{(2)}(\delta k)=1$ if there is no correlation in the system and $g^{(2)}(\delta k)\neq 1$ otherwise (see Fig.~\ref{Fig:Sup4}b).

To ensure a proper normalization we calculate $G_{\text{NC}}^{(2)}$ with as a procedure similar to that used for $G^{(2)}$ but by getting rid off the correlations before performing the calculation of the histogram. Removing the correlations is obtained from considering the atoms belonging to all the shots since there is no correlations between two atoms belonging to two different shots of the experiment. The value of $G_{\text{NC}}^{(2)}$ for the same set of data mentioned earlier is given in Fig.~\ref{Fig:Sup4}a. 

The result of dividing $G^{(2)}$ by $G_{\text{NC}}^{(2)}$, {\it i.e.}, $g^{(2)}(\delta k)=G^{(2)}(\delta k)/G_{\text{NC}}^{(2)}(\delta k)$ is plotted in Fig.~\ref{Fig:Sup4}b for the same set of data used in Fig.~\ref{Fig:Sup4}a.
\newline

\begin{center} 
\textit{Effect of transverse integration }
\end{center}

\begin{figure}[ht!]
\includegraphics[width=\columnwidth]{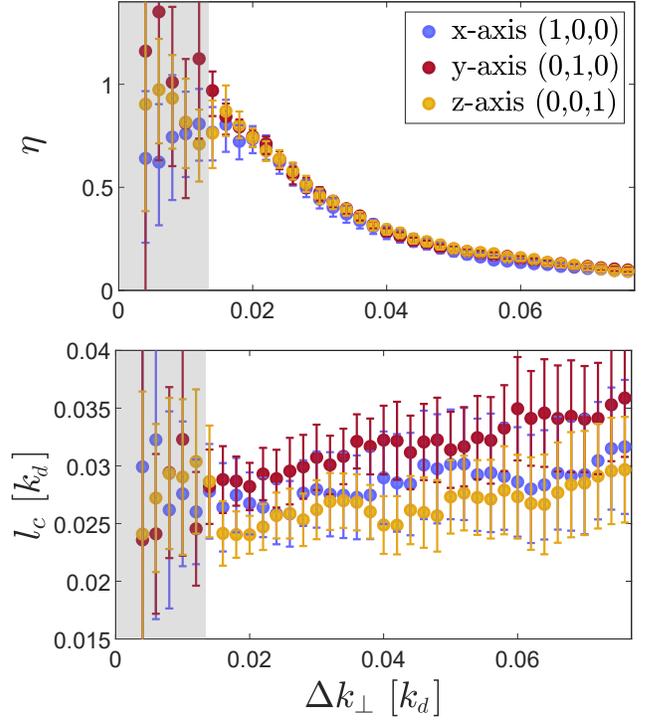}
\caption{\textbf{Effect of the transverse integration on $g^{(2)}(\delta k)$.} \textbf{a)} Amplitude of the bunching $\eta$ as a function of $\Delta k_{\perp}$ for the data used  in Fig.~\ref{Fig2}. \textbf{b)} Correlation length $l_c$ as a function of $\Delta k_{\perp}$. Grey areas on both graphs correspond to the region where the signal is too noisy to provide quantitative information.}
\label{Fig:Sup_int}
\end{figure}

The volume of integration $\Delta k\times \Delta k_\perp^2$ may affect the shape of $g^{(2)}(\delta k)$ when $\Delta k$, $\Delta k_\perp$ $\approx l_c$, with $l_{c}$ the correlation length of $g^{(2)}(\delta k)$ as defined in the main text. To avoid affecting the  size and shape of the correlation peaks, we use $\Delta k \approx 0.1l_c$. See Tab.~\ref{tablecorrelation}. Although the transverse integration does not modify the shape of the correlation peaks, it results in decreasing the bunching amplitude $\eta$ defined by $g^{(2)}(\delta k)=1+\eta \exp(-2 \delta k^2/l_{c}^2)$. Fig.~\ref{Fig:Sup_int} displays the measured value of $\eta$ as function of $\Delta k_{\perp}$, illustrating this assertion.

For the data shown  in Fig.~\ref{Fig2}, we use $\Delta k_{\perp}=0.015~k_d$, which is smaller than $l_c=0.027(4) ~k_d$. At lower transverse integration, the signal becomes too noisy (see Fig.~\ref{Fig:Sup_int}). To further increase the signal-to-noise ratio in order to reach a precise measurement of $l_c$, we use $\Delta k_{\perp}\approx l_c$  in Fig.~\ref{Fig3}. Consequently, the bunching amplitude $\eta $ lies between $0.3$ and $0.5$ (see Tab.~\ref{tablecorrelation}). 

\begin{table}[ht!]
\begin{tabular}{l c l c l c l c l c l c l }
 $L$  & $N$   & $\Delta N$  &  $l_c$ & $\Delta k_{\perp}$     &$\Delta k_{||}$ & \hspace{0.02cm} $\eta$\\ \hline
   [$d$] &  [$10^3$]  &  [$10^3$] &  $[k_d]$ & $ [k_d]$     &$[k_d]$  & \hspace{0.06cm} - \\ \hline
 23 & 3 & \hspace{0.06cm} 1 &0.038 & 0.03  &0.002  &0.38\\ \hline
 25&   5 & \hspace{0.06cm} 1&  0.035& 0.03 & 0.002 & 0.32\\ \hline
 30 &   12 & \hspace{0.06cm} 2 &0.03 & 0.03 & 0.001 & 0.51 \\ \hline
  34 &   16 & \hspace{0.06cm} 3 & 0.027& 0.03 & 0.001 & 0.43\\ \hline
   36 &   20 & \hspace{0.06cm} 4 & 0.025& 0.03 & 0.001 &  0.42\\ \hline
38 &  22 &  \hspace{0.06cm} 4 &0.023& 0.03 & 0.001 & 0.40\\
 \hline

\end{tabular}
\caption{Parameters used to compute the data of  Fig.~\ref{Fig3}. In order to measure the correlation length with a good precision, we use $\Delta k_\perp \approx l_c$, resulting in $\eta=g^{(2)}(\delta k =0)-1<1$.}
\label{tablecorrelation}
\end{table}

\begin{center} 
\textit{Numerical calculation of the correlation length $l_{c}$}
\end{center}

The dashed line of Fig.~\ref{Fig3} has been obtained numerically with the following procedure. Firstly, we calculate the three-dimensional distribution of atoms from the Gutzwiller ansatz which is valid at large $U/J$. Secondly, we assume that atoms in the different lattice sites are fully decoupled. The resulting density operator being Gaussian, we can use Wick theorem to show that $g^{(2)}(\delta k)=1 + |g^{(1)}(\delta k)|^2$ and thus evaluate numerically $g^{(2)}(\delta k)$ from summing the contributions of all the atoms to the one-body correlation $g^{(1)}$. This procedure is identical to that used in \cite{folling2005}. 
\newline

\begin{center} 
\textit{3D integral of the bunching peaks on the reciprocal lattice}
\end{center}

As explained in the main text, we find an identical 3D integral for the bunching peaks associated with the reciprocal lattice. The table \ref{tablevolcoh} contains the measured amplitudes and correlation lengths of the different peaks in $g^{(2)}$ shown in Fig.~\ref{Fig2}(b). We also give the values of the ratio between the 3D integrals of the peak located at $\delta k=0$ and that located at the position $\delta k \neq 0$ of the reciprocal lattice, $\frac{\eta l_{c} ^3}{\eta^{\neq 0} (l_{c}^{\neq 0})^3} $. 

\begin{table}[ht!]
\begin{tabular}{l c l c l c l c l  }
Direction  & [1,0,0]   & [1,1,0]  &  [1,1,1]  \\ \hline
 $\eta$ & 0.98(2) & 0.94(3) & 1.0(2) \\ \hline
$l_{c}$ &   0.029(1) & 0.031(1) &  0.026(1) \\ \hline
$\eta^{\neq 0} $ &   0.51(1) & 0.49(1) & 0.45(2) \\ \hline
$l_{c}^{\delta k \neq 0}$ &   0.035(1) &0.037(2) & 0.033(1)\\ \hline
$\frac{\eta}{\eta^{\neq 0}} \left ( \frac{l_{c}}{l_{c}^{\neq 0}} \right)^3$ &   1.04(6) & 1.15(8) & 1.12(8)\\
 \hline
\end{tabular}
\caption{Results from fitting the bunching peaks in $g^{(2)}(\delta k)$ with a Gaussian. $\eta$ and $l_{c}$ are respectively the amplitude and the correlation length of the bunching at $\delta k=0$. $\eta^{\neq 0}$ and $l_{c}^{\neq 0}$ are respectively the amplitude and the correlation length of the bunching at $\delta k\neq0$. The ratio of the 3D integral of the bunching peaks at $\delta k=0$ and at $\delta k \neq0$ is close to 1 along any direction of the reciprocal lattice.}
\label{tablevolcoh}
\end{table}

As the 3D integral of the bunching peak is found constant at any $\delta \vec{k}=\vec{K}$, the lower amplitude in $g^{(2)}(\delta \vec{k})$ observed at $\delta \vec{k}=\vec{K}\neq\vec{0}$ results from the measured larger correlation length. We stress that the measured increase in the correlation length amounts only to a few times $k_{a}/1000$. The capability to measure such a small difference demonstrates the outstanding performances of the He$^*$ detector. On the other hand, it is most probable that the origin of this tiny discrepancy lies in small imperfections of the He$^*$ detector. 

More specifically, a detector made of Micro-Channel Plates and a delay-line anode -- as in the He$^*$ detector -- is better at measuring small distances than large ones between individual particles. Small distortions in the image of a regular pattern were indeed reported on distances comparable with the MCP diameter \cite{jagutzki2002}. In addition, a recent investigation with MeV $\alpha$ particles has also shown that the uncertainty on the measure of distances of the order of the MCP diameter may be up to 4 times larger than that on the measure of small distances \cite{hong2016}. To our knowledge a full understanding of these distortions is missing but several origins for these imperfections have been identified. It includes the inhomogeneity of the electric field on the edges of the MCP and the presence of mechanical imperfections in the delay-line anode -- {\it e.g.} an imperfect winding of the cables realising the electronic wave-guide. 

In our experiment, the resolution with which the position of individual atoms is reconstructed is estimated to be $ \sigma \sim 2.5 \times 10^{-3} k_{d}$. The increase in the correlation length $l_{c}^{\neq 0} \simeq1.2 l_{c}$ is compatible with an accuracy to measure large particle separation which is 3 times worse  ($\sim 7.5 \times 10^{-3} k_{d}$).


\end{document}